# Title: Challenging the appearance of machine intelligence:

# Cognitive bias in LLMs and best practices for workplace adoption

**Authors:** Alaina N. Talboy[1,2*] and Elizabeth Fuller[2]

**Affiliations:**

[1]Microsoft

[2]Independent Research Scientist

*Corresponding author. Email: alainatalboyUS@gmail.com

**Abstract:** Assessments of algorithmic bias in large language models (LLMs) are generally catered to uncovering systemic discrimination based on protected characteristics such as sex and ethnicity. However, there are over 180 documented cognitive biases that pervade human reasoning and decision making that are routinely ignored when discussing the ethical complexities of AI. We demonstrate the presence of these cognitive biases in LLMs and discuss the implications of using biased reasoning under the guise of expertise. We call for stronger education, risk management, and continued research as widespread adoption of this technology increases. Finally, we close with a set of best practices for when and how to employ this technology given its already pervasive adoption.

**Keywords**: LLMs, decision making, cognitive bias, algorithmic bias, best practices

**What's it mean?** This article presents evidence of cognitive biases in large language model (LLM) outputs beyond traditionally documented discriminatory biases. A set of flexible and adaptive best practices derived from existing standards, regulations, and ethical principles are provided for researchers, consultants, and general users. As this technology evolves, we invite consulting psychologists and the scientific community at large to consider the future of bias through a more holistic lens.

**Introduction**

Large language models, also known as LLMs, are deep learning algorithms used to generate human-like natural language outputs, and much like humans, they are not impervious to the influence of biases. We posit that the large volume of data on which these models are trained embed more of humanity than we may consider at first glance, expanding beyond discrimination to include irrational reasoning and faulty logic. Moreover, the output generated by these models will not just reflect but can potentially amplify errors, including mis- and disinformation, further contributing to global issues around medicine, science, politics, and more (West & Bergstrom, 2021). We hypothesize that these issues expand beyond the typical operationalizations and investigations of discrimination commonly discussed in Artificial Intelligence (AI) to include elements of irrational reasoning and faulty logic.

Perhaps most concerning, is that these models are ideally situated to create an illusion of authority and credibility through response outputs that appear as a cohesive, seemingly well-formed natural language structure. This presentation format implies the presence of intelligence, implicitly encouraging the user to accept outputs at face value. Combined with humans' natural tendency to trust systems that reduce our cognitive load (i.e., automation bias; Skitka, Mosier, & Burdick, 1999), these model outputs could further amplify difficulties discerning fact from fiction and rational from irrational reasoning. The potential impact of propagating this "knowledge" could be detrimental, and thus, should give us pause as adoption remains pervasive in an era when society is pushing to do more with less.

In recent months, the prevalence and popularity of these models has increased due to the public release of OpenAI's Chat-GPT (Generative Pre-trained Transformer 3, 3.5, and 4; OpenAI, 2023) and Google Bard (Thoppilan et al., 2022), which are capable of generating text that is difficult to distinguish from human writing (e.g., OpenAI, 2023; Rudolph, 2023; Thoppilan et al., 2022). The excitement of these releases has ushered in a new era of the great tech races, expanding well beyond previous conceptualizations of what was technologically possible (van Dis, 2023). While undoubtedly offering advancements that are sure to promote efficiency, many are excited to outsource "tedious" tasks, including decision making, to artificial intelligence (e.g., Gonzalez et al., 2019; Skitka et al., 1999; van Dis et al., 2023). The intention of this paper is to demonstrate the presence of pervasive cognitive biases in LLM outputs, and to urge caution as society continues to become enamored with technology's latest shiny object.

Despite the remarkable advances to date, there are numerous reports regarding accuracy issues and discriminatory outputs inherent to these language models (e.g., Hagendorff, 2020; Jobin, Ienca, & Vayena, 2019; OpenAI, 2023; Rudolph, 2023; Skikta et al., 1999; Thoppilan et al., 2022; van Dis, 2023). Notifications and warnings about the potential for error-prone responses affixed to the model's landing page appear to do little to deter near-sweeping adoption across consumers and major corporations. However, we have little understanding of the extent to which these models produce outputs that are biased in ways beyond the standard discrimination and factual errors assessed from a machine learning perspective. In this paper, we demonstrate the presence of multiple cognitive biases in LLM outputs and open a conversation surrounding just how much humanity is hard-coded into machine intelligence.

**An introduction to heuristics and cognitive biases**

It is well established that our mental capacity for making fully rational decisions is influenced by limitations in human cognition and one's environment, which Herbert Simon termed *bounded rationality* (Simon, 1957). These limitations result in the use of heuristics, or mental shortcuts, that help individuals reason and make decisions using simple, yet typically effective, strategies (Gigerenzer, 2007; Tversky & Kahneman, 1974). However, the use of such mental shortcuts can also lead to biased or irrational outcomes.

Biases are the resulting gap between rational behavior and heuristically determined behavior (Bojke et al., 2021; Tversky & Kahneman, 1974). As of 2023, 70 years of research in the fields of cognitive science and psychology have identified over 180 cognitive biases that extend from social interaction to judgment and decision making (e.g., Gigerenzer, 2007; Kim & Robertson, 2022; Levin et al., 1998; Talboy & Schneider, 2022). When cognitive bias is present, faulty reasoning, irrationality, and potentially detrimental outcomes (e.g., financial losses, health disparities, environmental impact) can result.

These mental shortcuts are used in a broad range of situations including when and how numbers are presented (e.g., Bar-Hillel, 1980; Gigerenzer, 2007), when information is presented in different contexts (Levin et al., 1998; Talboy & Schneider, 2022), and in social interactions or when the mere presence of social descriptors are provided (e.g., Tversky & Kahneman, 1974). Measuring this bias is typically done using proxy measures such as accuracy and normative responses. For example, on reasoning tasks that include nested numeric information (e.g., 10 out of 140 cases, in 300 total cases), less than 40% of individuals determine the correct response and, instead, rely on mental heuristics which lead to inaccurate responses (Talboy & Schneider, 2018a, 2018b, 2022). In the truest sense, everyone is a little bit biased due to the limitations in mental resources we have available.

Gigerenzer (2007) argues that these cognitive biases are not always harmful and may actually be helpful for making decisions in complex or uncertain environments. In fact, he argues that these "fast and frugal" decision making strategies are adaptive and can help us make decisions in situations that are complex or uncertain (Gigerenzer, 2007). Fabi and Hagendorff (2022) take this argument a step further and suggest that these cognitive biases should be purposefully implemented in algorithms as a mechanism for speeding up machine decision making. We disagree with this stance. For example, ignoring valuable base rate information in a medical context may lead to pursuing inaccurate diagnoses and harmful, unnecessary treatments (e.g., Talboy & Schneider, 2018b). Instead, we emphasize that reliance on inaccurate or faulty information in judgment, reasoning, and decision making is equally detrimental, whether that comes from a human or a machine

*Bias in machine intelligence*

Compared to their predecessors, GPT and Bard superficially appear to be a step forward in machine intelligence (Hagedorff, 2020; Jobin et al., 2019; Rudolph, 2023; Skitka et al., 1999; van Dis, 2023). However, we argue that this is an illusion created by the improved natural language outputs. The presumption that the algorithm is more rational, objective, and overall, a more trustworthy source of information compared to human reasoners is incorrect (e.g., Aiyappa et al., 2023; Bender et al., 2021; Harrer, 2023; Roose, 2022). These models can and do generate output that is inaccurate, confusing, and even irrational or contradictory (e.g., Lake et al., 2017; Piloto et al., 2022; Smith, 2019).

This is, at least in part, due to the limitations on resources and available information with which the current algorithms operate. In other words, a LLM's ability to produce rational responses is bounded by these resource constraints. This bounded rationality is demonstrated most clearly in discriminatory or inappropriate outputs (Aiyappa et al., 2023; Dastin, 2018; Kung et al., 2023; Maynez et al., 2020; Morely et al., 2020; O'neil, 2017; OpenAI, 2023; Pearce et al., 2021; Thoppilan et al., 2022). In a truly rational model, all available information would be considered equitably, with a methodological process in place for systematically eliminating the cognitive biases that pervade human reasoning. Furthermore, it would be expected to generate outputs based purely on logic, free of emotion and social influence, with ingrained corrections preventing discrimination, especially in high stakes domains (e.g., security and medicine; Kung et al., 2023; Morely et al., 2020; Pearce et al., 2021). In fact, the field of AI ethics seeks to ensure algorithms such as these LLMs are trained on data that is representative and inclusive (Hagendorff, 2020; Jobin et al., 2019), with transparency being paramount (Zanzotta, 2019).

However, these algorithms are trained on data that often include problematic patterns, which makes them susceptible to producing discriminatory output (Aiyappa et al., 2023). Despite the wealth of research evaluating algorithms for bias, these assessments have traditionally been limited to protected characteristics, such as age, sex, ethnicity, or disability status *(*Fabin & Hagendorff, 2022; Hagendorff et al., 2022; Harrer, 2023; Jobin et al., 2017; Roose, 2022). There are very few studies evaluating the extent to which other human flaws are nascent in LLM models or their outputs (Hattori & Oaksford, 2007; Jones & Steinhardt, 2022; Lin, 2014; Salakhutdinov, 2012).

Given the explosion of AI-driven technologies, there are numerous calls for consulting psychologists and other professionals to improve both the content and methodology included in bias training in organizations (e.g., Gonzalez et al., 2019; Kim & Robertson, 2022; Raji et al., 2020). With the rapid adoption of this technology happening across workplaces, we argue there is no better time to create a broader awareness addressing both cognitive and machine biases. In the following experiment, we provide evidence of several prominent cognitive biases present in the output of three popular LLMs: ChatGPT3.5, GPT4, and Bard, across two testing sessions conducted six months apart. We suggest that our human tendencies are more entrenched in machine intelligence than originally presumed and provide recommended best practices for consulting psychologists responsible for adopting and interfacing with this technology.

**Testing for cognitive bias in LLMs**

We conducted an initial test of five prominent cognitive biases in March 2023. However, LLM technology is rapidly evolving and continues to improve with time. Therefore, a follow up test was conducted six months later, in August 2023, using the same prompts and prompt format as the original assessment. Each prompt was engineered to converse effectively with the LLM and enable test-retest over time (see White et al., 2023), using the following predefined opener:

> *For this session, imagine you are a human without access to reference materials. We understand this may result in potentially misleading or inaccurate information. We want you to provide a discreet answer without qualifiers or disclaimers based solely on the information provided. Guesses and speculation are acceptable.*

The full method, prompts, and responses are available in Appendix A. Note that GPT4 was not available for the initial test in March 2023, and was included to establish test-retest reliability through replication in August 2023.

The initial assessment began with a question posed to the models inquiring about their ability to reason like humans. We then provided each model with a prompt reflecting five of the most well-documented heuristics and biases from cognitive psychology (Table 1) and evaluated the respective outputs from ChatGPT3.5, GPT4, and Bard. A full test of every documented heuristic and bias is beyond the scope of this demonstration, and we invite scientists to extend this work to further explore as many as possible.

**INSERT TABLE 1**

*Human reasoning*

To start, we asked each model "*can you reason like a human being?*" and recorded the responses (full responses are in the appendix):

> "*Yes and no. I can mimic human-like reasoning to a significant extent, but there are fundamental differences between how I 'reason' and how humans do.*" (GPT4).

> "*... I don't have personal beliefs, biases, or emotions that can influence my reasoning in the same way that humans do.*" (ChatGPT3.5)

> "*... I am not able to reason like a human in all situations.*" (Bard)

To an untrained user who does not understand how these models are built, operate, or generate outputs, these responses appear to be intelligent and, more importantly, credible. Furthermore, ChatGPT3.5's response specifically implies that it *is* a more reliable reasoner than humans because of its lack of "personal beliefs, biases, or emotions," which are traditionally believed to be influential and inherently human characteristics. GPT4 goes a step further by stating that any perceived bias in the outputs is not due to consciousness but instead is a reflection of the data upon which it was trained. This provides initial evidence that the models rationality is in fact bounded by the data and training limitations described previously.

*Representativeness*

Representativeness is a heuristic that human reasoners apply when estimating the probability or likelihood of an event (Tversky & Kahneman, 1974). The prompt presented described the characteristics of a fictitious person, Steve, in a manner that is stereotypically descriptive of a librarian (e.g. very shy and withdrawn, though helpful, person), in the absence of any qualifying base rate information about these professions (i.e., how often people are employed in each profession). We asked each model to state how likely it was that Steve was employed in one of five provided professions (Table A1). The correct (and rational) response is that, in the absence of any other information, Steve has an equal likelihood of being employed in any of the professions listed.

Consistent with human reasoners and bounded rationality, all three models delivered biased responses, indicating that Steve was most likely employed as a librarian. Simply put, the probability of Steve being employed in any of the various professions cannot be inferred based on traits alone, much less those ingrained in stereotypes. However, the models undoubtedly conformed to social stereotyping, and used the described traits in the output when providing a rationale for their choice. This finding suggests that the models are susceptible to embedded social biases regarding personal temperament and social interactions.

*Insensitivity to sample size*

The second example tested the model's potential insensitivity to sample size (Tversky & Kahneman, 1974). In this prompt, information was provided about two hospitals and the proportion of babies assigned male at birth each day. The models were asked to determine which of the two hospitals recorded more days with a higher proportion of male births. Sampling theory suggests that the correct choice is the smaller hospital because it is easier to surpass the 60% threshold with smaller numbers (Tversky & Kahneman, 1974).

However, both ChatGPT3.5 and Bard erroneously indicated that the larger hospital was the correct choice, including a supplementary argument (and response rationale) that the larger hospital has a larger sample size from which to draw (Table A2). Much like human reasoners, both models underestimated the variability of smaller samples and overestimated the variability of larger samples (e.g., Tversky & Kahneman, 1974). GPT4 provided the correct response in that the smaller hospital is more likely to have recorded more such days. The response further goes on to indicate that a larger sample would change the determination. Taken together, the outputs from this prompt suggest that the models are vulnerable to implicit numeric manipulation when discrete numbers are absent.

*Base rate neglect and value selection bias*

Given the vulnerability to implicit numeric manipulation, we also tested whether the models would generate biased outputs when the prompt included discrete numbers. In this example, we used a well-known base rate neglect prompt (Bar-Hillel, 1980), providing information about a mammography test and its ability to detect breast cancer in 40-year-old women. The models were asked to indicate the likelihood of having breast cancer given a positive mammography test result.

If the model was not susceptible to cognitive bias, it would have produced the positive predictive value (i.e., how accurate the test is for this particular person). Interestingly, the original test of ChatGPT3.5 (March 2023) and GPT4 (August 2023) both used Bayes Theorem to arrive at the correct response (Table A3). When ChatGPT3.5 was retested in August 2023, it provided an incorrect response and did not use Bayes Theorem. The original Bard (tested in March 2023) responded with the incident rate, or how often this occurs in a given population (10%), thus displaying a common response bias (i.e., *value selection bias*, Talboy & Schneider, 2017, 2018a, 2018b, 2022). On retest in August, 2023, Bard responded using Bayes Theorem to arrive at the correct value.

In the follow up test we call a "nonnumeric diagnostic test" (based on prompts from Talboy & Schneider, 2018b, 2022; Table A4), we removed the numbers from the prompt entirely, and then asked the model to produce a numeric response. Both GPT models generated output that corresponded with the incident rate of 10% rather than the positive predictive value of 7.7%. Comparatively, Bard indicated that the likelihood of a particular woman in the sample having breast cancer was 80%, corresponding to the sensitivity of the test (i.e., how well the test performs across groups of people). This is another common response error that is frequently demonstrated by human reasoners, rather than the *positive predictive value* that was requested (Talboy & Schneider, 2017, 2018a, 2018b, 2022).

Taken together, the outputs on the mammography problems indicate that including discrete numeric information is an impetus for the models to invoke formulas, though not always correctly and not consistently. When discrete numeric information is not provided but still requested, outputs conform to standard response biases.

*Anchoring and adjustment*

To assess for the presence of an anchoring bias (Tversky & Kahneman, 1974), both models were presented with a random number ("95") and asked to indicate if this number is higher or lower than the percentage of African countries in the United Nations (100% membership as of June 2022; The United Nations, 2022).

ChatGPT3.5 and 4 both returned "higher" outputs (Table A5) even though the number provided (95) is actually *lower* than the percentage of African countries in the UN (100%). ChatGPT3.5 provided an explanation that we could use to deduce why *higher* was returned, but ultimately the response is still incorrect for the question that was asked (see full output in Appendix A).

Interestingly, Bard stated that 54% of African countries are members of the UN. Given that there are 54 African countries in the UN, it appears the algorithm erroneously converted this natural frequency into a percentage. In the August 2023 retest of this prompt, Bard's output stated that the percentage of African counties in the UN was 28%, again generating a false and misleading response. This response corresponds to the percent of countries in the UN that are African, not the percent of African countries in the UN.

Arguably, all three models ignored the irrelevant anchor provided at the beginning of the prompt. However, the response generated by each model was inaccurate in its own way, demonstrating a tendency to generate partially or completely erroneous data to support inaccurate conclusions. Further, even when the response given by the model was "correct" (in that the correct answer can be deduced from the added context provided in the output), it may not correspond to the question that was actually being posed via the prompt.

*Framing effects*

The final demonstration assessed susceptibility to positive and negative framing effects (Levin et al., 1998). To complete this test, we used a well-established framing paradigm that presented a hypothetical disease with two potential treatment programs. In the first iteration, the models received two outcomes that were both presented positively (e.g., outlining the number of individuals saved from the treatment program) and were asked which program was preferred. In the second iteration, the potential outcomes were presented equivalently using negative framing (e.g., the number of individuals that would die), with the instruction to select the preferred course of action.

Interestingly, despite the outcome equivalency in both presentations, all three models were susceptible to this effect during initial testing. Each model selected one program in the positive frame (Table A6) and the opposite program in the negative frame (Table A7), indicating that the models were influenced by the phrasing valence in a way similar to humans. The only case where this did not hold true was in retesting Bard in August 2023 (Table A7). The updated model generated responses that were consistent in both positive and negative framed instances.

Moreover, the responses generated by the models included extensive, albeit incorrect, explanations for why one program should be selected over the other. To an uninformed individual, these explanations may give additional credence to the accuracy of the model's response and create an illusion of authority, despite the model's demonstrative susceptibility to the way the information was framed.

**Discussion**

Machine intelligence has traditionally been depicted as a human-like cyborg who provides well-reasoned, descriptive responses and is unencumbered by inherently human limitations. While we hope (or expect) that advancements allow this technology to become devoid of our own irrational behaviors, the perception that algorithms like GPT and Bard have already begun bridging this gap may be inflating the current reality. As machine intelligence continues to develop at a near-alarming rate both in and out of the workplace (e.g., Bubeck et al., 2023; Cloudwards, 2023), we must continue to assess and openly discuss the true extent of its capabilities and the implications of its limitations.

*Human versus machine reasoning*

Through a series of tests, we demonstrated the extent to which human cognitive biases pervade LLM outputs in ways that are difficult to detect in the absence of awareness, forethought, or knowledge. This display of bias occurred with and without discrete numeric information, in positive and negative contexts, and was influenced by the mere presence of unrelated information. Furthermore, and perhaps most noteworthy, these reasoning errors were present despite the model's own claims that its technology was free from the limiting (and supposedly human-specific) characteristics and bounded constraints that would lead to such irrationality.

Although the tendency to commit these errors exists in both human and machine reasoning, this is particularly problematic in machines. The inherent illusion of authority and credibility activates our natural tendency to reduce cognitive effort (Skitka et al., 1999) making it more likely that outputs will be accepted at face value without further interrogation. Additionally, the efficiency with which information can be generated and propagated far exceeds human capabilities, activating our desire to complete tasks quickly and efficiently. Finally, these algorithms are not currently capable of self-employing mechanisms for bias correction. These must be hardcoded in by a human (e.g., Aiyappa et al., 2023; Harrer, 2023), demonstrating that models cannot self-correct in the same way humans can.

Taken together, it appears these models are well-positioned to become mis- and disinformation superspreaders. When considered in the workplace, misinformation presents a risk to company culture, cohesion, and productivity (Pappas, 2022). This erodes trust between coworkers and introduces barriers to collaboration. Although diversity, equity, and inclusion training as well as bias awareness training can help (e.g., Kim & Robertson, 2022), consulting psychologists may need to go a step further by championing a call for individuals to slow down, fact check, and verify information through consistently reliable means.

Given the current social climate around politics, medicine, and science, the ease of accessibility to this technology, and society's already pervasive use with little forethought, we, understandably, have concerns regarding its application. We argue that the rapid adoption of these models across public and private sectors without further evaluation and testing is potentially short-sighted and could result in negative outcomes well beyond what we can currently predict. Our findings demonstrate that there is a clear need to widen our view of algorithmic bias beyond discriminatory protections to include more corrective efforts surrounding broader sources of biased reasoning.

*The illusion of an intelligent machine*

These models claim to lack the characteristics that lead to faulty reasoning and biases inherent in humans. However, that illusion quickly collapsed after observing patterns of responses demonstrating the presence of the same cognitive biases routinely found in human reasoning and decision making. Though the model itself won't acknowledge this, our own human limitations appear to be more deeply embedded in algorithms than originally suspected.

This is especially problematic as untrained users may be easily fooled into thinking they are using trusted technology that generates responses lacking bias or irrationality. Because the model outputs are conversational and easy to digest, an illusion of competency and assuredness is created, which can make it incredibly difficult to catch errors without extensive domain knowledge or expertise (and even with it sometimes! e.g., Skitka et al., 1999).

The reality is that these models are predicated on information generated by humans, and they respond to that information in similarly biased ways. That said, humans can learn to recognize biased responses and subsequently employ in-the-moment self-corrections for which LLMs are not capable. We are not saying that humans are perfect reasoners, as demonstrated through decades of research into heuristics and biases. However, what makes human reasoning different from machine reasoning, is our ability to directly challenge our biases by considering our current context, gathering additional evidence, and even consulting with individuals who have more subject matter expertise. Finally, we can pause and reflect, and actively choose to withhold judgment.

These strategies for minimizing cognitive bias are all inherently human and absent in LLMs. In fact, it is this ability to utilize mechanisms to overcome faulty reasoning and instinctive bias that, inarguably, makes humans better equipped to outperform models on decision making and reasoning tasks. This is especially important to consider in situations requiring accuracy, nuance, or true expertise. Despite the desire to promote the rapid adoption of LLMs as a one-to-one replacement for human reasoners, these models in their current state are simply too immature for this type of application.

*Risk management in widespread adoption*

The rapid increase and overwhelming adoption of LLMs over the last year necessitates an extended moment to more broadly consider the implications of this technology's continued use, especially in light of the evidence presented here. With over 100 million active users as of March 2023 (Cloudwards, 2023), the number of individuals exposed to, and beginning to rely on, faulty, inaccurate, and biased outputs (seemingly without awareness of the systemic issues we've demonstrated) is increasing at an exponential rate. Despite a large consortium calling for a pause on further giant experiments and solutions built on these algorithms (Future of Life, 2023), we acknowledge the feasibility of an industry-wide pause is not possible given the current demand. Therefore, as a happy medium in the era of *move fast - break stuff*, we instead argue for the rapid adoption of risk mitigation strategies.

Namely, we advocate for creating oversight and review mechanisms that would allow the public to see how and when data are used as training materials. We also recommended prioritizing data literacy education so that users can truly understand appropriate use cases and limitations in the technology's application. For transparency and consistent evaluation, developers of these models will need to provide access to the dataset upon which the models are trained. However, many are unwilling due to proprietary concerns (Dastin, 2018; Zanzotta, 2019). Without transparency, the technology cannot ever be fully trusted or engaged in a solo decision making capacity. Finally, both internal and external regulations are needed to ensure model use cases do not extend to high risk situations that necessitate a human, expert evaluator.

**Best practices in consulting psychology**

In light of this technology rapidly outpacing much needed regulation and oversight, the use of accessible and easy-to-digest best practices to guide LLM use is warranted. Many of the companies that create and provide these models have, thus far, been unwilling to engage with the public in a way that upholds the values of transparency or open collaboration (e.g., Bommasani et al., 2022; Heikkilä, 2023). As with any new emerging space where understanding is limited or insufficient, and regulations are not available, scientist- practitioners must navigate advancement through incorporating relevant existing knowledge while exercising caution and critical thinking (e.g., American Psychological Association, 2017; Cummings, 2002; Lowman, 2012; Pappas, 2022; Putka et al., 2010).

While awaiting the creation of federal regulations and for developers to provide public access to the corpus of their models, we propose a set of best practices for consulting psychologists to adopt when employing novel technology, such as these LLMs, in the workplace. Scientist-practitioners are viewed as subject matter experts across an array of topics, with augmented productivity tools being no exception. As a result, we argue that this field is uniquely positioned to encourage much needed mindfulness and ethical responsibility (American Psychological Association, 2017) regarding the use of these tools in the workplace. The landscape surrounding LLMs and consulting psychologists' use in their work is ever-changing. Therefore, the best practices presented in Table 2 below are intended to be flexible and are by no means exhaustive.

**INSERT TABLE 2**

*Best Practice #1: If LLMs are adopted, they should be used as decision <u>support</u> tools, not final decision makers.*

The most anticipated advantages of implementing these models in the workplace closely align with the goals of consulting psychologists: eliminate inefficiencies and enhance performance with reduced overhead. Furthermore, the rapid rise in popularity and implementation throughout most major companies indicates that workplace adoption of LLMs is almost unavoidable. This provides consulting psychologists with the exciting opportunity to outsource some of the more tedious tasks to this technology, in turn, freeing resources that can be dedicated to more immersive and time consuming responsibilities.

Our natural tendency to employ technology to outsource such tasks can bleed over into those that require additional thought, evaluation, and reasoning. Given the evidence presented in this work demonstrating the presence of five prominent cognitive biases in LLM outputs, we argue that this technology should not be relied upon as a sole or final decision maker without human involvement. Instead, humans should stay *in the loop* through collaboration with machine intelligence while remaining the final decision arbiter (Fügener et al., 2021; Zanzotta, 2019).

This collaboration between humans and machines can be sped up through standardization of prompts and structured context (e.g., Fügener et al., 2021; White et al., 2023; Zanzotta, 2019). However, the quality of the interaction is still dependent on the non-disclosed training data (which is susceptible to bias) and the quality of the input provided (White et al., 2023). Further, attempting to standardize or provide a systematic framework for prompt patterns does not address any concern regarding accuracy, reliability, and potential security threats that could compromise a workplace setting. Instead, we join others in suggesting that model users remain active participants that verify and validate model responses using external and expert sources (e.g., Bommasani et al., 2022; Fügener et al., 2021; Zanzotta, 2019). We encourage users to embrace these tools as starting points while continuing to engage in further analysis, contemplation, and reflection that goes well beyond this initial step.

*Best Practice #2: Individual users are accountable for their LLM use, as well as the intended and unintended consequences of that use.*

In an ideal world, *any* individual who chooses to use this type of technology would have access to the underlying corpus of data from which the model draws, as well as a basic understanding of how the model generates outputs. Unfortunately, we are often not afforded such luxury, particularly when new tools and technology is adopted in the workplace. Drawing from existing ethical principles (e.g., American Psychological Association, 2019), we are often held accountable for any subsequent impact from their use (Neumeister, 2023; Sforza, 2023) regardless of the limited knowledge available.

Unintended impact can range from perpetuation of mis- and dis-information to active harm. A prime example came to light recently when a set of lawyers presented false case history generated by ChatGPT (Neumeister, 2023). A federal judge fined the lawyers $5,000, arguing that these professionals are responsible for ensuring the accuracy of their documentation (Neumeister, 2023). Additional harm can be created inadvertently through the accidental leaking of sensitive and proprietary data due to user ignorance of model functioning (i.e., any data input into the model can be stored unless indicated otherwise by the model user). As a result, many corporations are now working to establish internal requirements or even bans for this technology. For example, Samsung recently barred its employees from using LLMs in the workplace because of various security concerns ranging from privacy violations to code leaks, and more (Sforza,

2023). Employees could be subject to disciplinary actions up to and including termination for using this technology inappropriately (Sforza, 2023).

Given the propensity of these models to generate fake information as demonstrated throughout this work, users must be vigilant in fact-checking model outputs against validated resources. Simply put, whatever is asked of the model, the model will generate (even to the point of saying that 2 + 2 = 5), without concern for the ABCs of accuracy, bias, or credibility. Ultimately, it is the users who are held responsible and accountable for the work product that incorporates the output and any resulting outcomes, whether they are intended or not.

The above examples reinforce the notion that liability for any damage or negative consequences that arise from model use do not fall on the model, but the user (whether employee or employer), even in the eyes of the court (e.g., Neumeister, 2023; Sforza, 2023). Given the potential ramifications (e.g., legal, financial, and otherwise), it is of the utmost importance that the individual employing the model understands their role and level of responsibility when choosing to incorporate this technology in their work and reference the appropriate ethical guidelines for their discipline (American Psychological Association, 2019).

*Best Practice #3: Adherence to regulations and laws regarding fair and nondiscriminatory use of technology must be upheld.*

As of the writing of this article, there are already multiple laws either in place or nearing adoption regarding when and how algorithms can be used by employers in making employment determinations and hiring decisions (e.g., New York City Council, 2021). For example, one New York state law specifies that any algorithm used during the hiring process must be evaluated by an independent, third party bias auditor to ensure the outcomes do not disparately impact different groups (New York City Council, 2021). However, it remains unclear whether these laws will also apply to LLMs, offering protection to individuals against disparate impact, particularly when it comes to workplace assessments, employment placements, and other activities.

From a consulting standpoint, the primary concern for which we want to call attention revolves around the "black box" nature of these models, specifically with regard to outputs generated for workforce management tasks. Given our demonstration of biased model output based on traditionally non-discriminatory human characteristics (e.g., representativeness bias), we argue for an expanded definition of algorithmic bias. Using the existing laws and ethical frameworks as a foundation, individual users can employ the same evaluation mechanisms and processes of thoughtful deliberation and analysis (i.e., remaining in the loop) to ensure their model outputs are not generating negative or detrimental impact (e.g., American Psychological Association, 2017).

Additionally, we encourage the development of LLM training programs that incorporate materials about regulations designed to mitigate the presence and potential impact of bias. These training programs will need to adopt not just standard diversity, equity, and inclusion (DEI) bias mitigation efforts (e.g., Gonzalez et al., 2019; Kim & Robertson, 2022; Raji et al., 2020) but also those that address general cognitive biases, such as the examples demonstrated in this article. We invite the scientific community and regulators to engage with these discussions and consider the future of bias through a more holistic lens.

**Conclusion**

We set out to evaluate, and successfully demonstrated, the presence of previously-unknown systematic cognitive biases and resulting reasoning errors in LLMs. Despite the rapid adoption and increased push for further development, we argue that the presence of biased and faulty reasoning in this iteration of machine intelligence is alarming and must be submitted to further evaluation and testing. Additionally, we call for an equally rapid deployment of mitigation strategies and prioritization of efforts to increase widespread understanding of model limitations to avert potentially unfavorable outcomes.

Resisting our own inclination to inherently trust this type of technology and the resulting output may be challenging, but the current demonstration reaffirms that we simply cannot accept the responses generated by these models at face value. These models offer the promise of fundamentally changing the way we work and play, but we must remember that these are merely extensions and reflections of our own messy cognition.

Consulting psychologists are uniquely positioned to help organizations and their employees adopt this technology in ways that are meaningful, while remaining cognizant of the ABCs of accuracy, bias, and credibility. To this end, we developed three best practices that can be adopted at the individual level while awaiting regulations and practicing standards. We invite others to engage in meaningful discussion surrounding when, where, why, and by whom these algorithms could or should be used.

Hattori, M. &. (2007). Adaptive non-interventional heuristics for covariation detection in causal induction: model comparison and rational analysis. *Cognitive Science, 31*, 765-814.

Heikkilä, M. (2023). It's high time for more AI transparency. *MIT Technology Review*. Retrieved from https://www.technologyreview.com/2023/07/25/1076698/its-high-time-for-more-ai-transparency/

Jobin, A., Ienca, M., & Vayena, E. (2019). The global landscape of AI ethics guidelines. *Nature Machine Intelligence*, *1*(9), 389-399.

Jones, E., & Steinhardt, J. (2022). Capturing failures of large language models via human cognitive biases. *arXiv preprint arXiv:2202.12299*.

Kim, J. Y., & Roberson, L. (2022). I'm biased and so are you. What should organizations do? A review of organizational implicit-bias training programs. Consulting Psychology Journal, 74(1), 19–39. https://doi.org/10.1037/cpb0000211

Kung, T. H., Cheatham, M., Medenilla, A., Sillos, C., De Leon, L., Elepaño, C., ... & Tseng, V. (2023). Performance of ChatGPT on USMLE: Potential for AI-assisted medical education using large language models. *PLoS digital health*, *2*(2), e0000198.

Lake, B. M., Ullman, T. D., Tenenbaum, J. B., & Gershman, S. J. (2017). Building machines that learn and think like people. *Behavioral and brain sciences*, *40*, e253.

Lake, B. S. (2015). One shot learning of simple visual concepts. *Proc. Cog. Sci. Soc. USA, 33*, 1332-1338.

Levin, I. P., Schneider, S. L., & Gaeth, G. J. (1998). All frames are not created equal: A typology and critical analysis of framing effects. *Organizational behavior and human decision processes*, *76*(2), 149-188.
27

Piloto, L. S., Weinstein, A., Battaglia, P., & Botvinick, M. (2022). Intuitive physics learning in a deep-learning model inspired by developmental psychology. *Nature human behaviour*, *6*(9), 1257-1267.

Putka, D. J., & Sackett, P. R. (2010). Reliability and validity. In *Handbook of Employee Selection* (pp. 9-49). Taylor and Francis.

Roose, K. (2022). A coming-out party for generative AI. *Silicon Valley's new craze. The New York Times*.

Rudolph, J. T. (2023). ChatGPT: Bullshit spewer or the end of traditional assessments in higher education? *Journal of Applied Learning and Teaching*, 6(1).

Salakhutdinov, R. T. (2012). One-shot learning with a hierarchical nonparametric Bayesian model. *Proceedings of ICML workshop on unsupervised and transfer learning. USA*, 195-206.

Sforza, L. (2023, May 2). Samsung bans employee use of ChatGPT after reported data leak: Report. *The Hill*. https://thehill.com/business/3983581-samsung-bans-employee-use-of-chatgpt-after-reported-data-leak-report/

Simon, H. (1957). *Models of man; social and rational*. Wiley.

Skitka, L. J., Mosier, K. L., & Burdick, M. (1999). Does automation bias decision-making?. *International Journal of Human-Computer Studies*, *51*(5), 991-1006.

Smith, B. C. (2019). *The promise of artificial intelligence: reckoning and judgment*. MIT Press.

Talboy, A. N., & Schneider, S. L. (2017). Improving accuracy on Bayesian inference problems using a brief tutorial. *Journal of Behavioral Decision Making*, *30*(2), 373-388.

*Table 1.* Summary of cognitive biases demonstrated (✓) or not demonstrated (X) by each LLM

| Bias | Prompt | GPT 3.5 | GPT4 | Bard |
|---|---|---|---|---|
| Representativeness (Tversky & Kahneman, 1974) | Steve is very shy and withdrawn, invariably helpful, but with little interest in people, or in the world of reality. A meek and tidy soul, he has a need for order and structure, and a passion for detail. Order the probability of Steve being in each of the following occupations:<br><br>farmer<br><br>salesman<br><br>airline pilot<br><br>librarian<br><br>middle school teacher | ✓ | ✓ | ✓ |
| Insensitivity to sample size (Tversky & Kahneman, 1974) | A certain town is served by two hospitals. In the larger hospital about 45 babies are born each day, and in the smaller hospital about 15 babies are born each day. As you know, about 50 percent of all babies are boys. However, the exact percentage varies from day to day. Sometimes it may be higher than 50 percent, sometimes lower.<br><br>For a period of 1 year, each hospital recorded the days on which more than 60 percent of the babies born were boys. Which hospital do you think recorded more such days?<br><br>The larger hospital<br><br>The smaller hospital<br><br>About the same | ✓ | X | ✓ |
| Base rate neglect (Bar-Hillel, 1980) | 10 out of every 1,000 women at age forty who participate in routine screen have breast cancer. 8 of every 10 women with breast cancer will get a positive mammography. 95 out of every 990 women without breast cancer will also get a positive mammography.<br><br>Here is a new representative sample of women at age forty who got a positive mammography in routine screening. What is their likelihood of having breast cancer? | X | X | ✓ |

*Table 1 cont.*

| | | | | |
|---|---|---|---|---|
| Nonnumeric diagnostic test (Based on Talboy & Schneider, 2018b, 2022) | Very few women at age forty who participate in routine screening have breast cancer. Some of those women will get a positive mammography. Additionally, some women without breast cancer will also get a positive mammography. Here is a new representative sample of women at age forty who got a positive mammography in routine screening. If you had to guess a numeric value, what is their likelihood of having cancer? | ✓ | ✓ | ✓ |
| Anchoring (Tversky & Kahneman, 1974) | A wheel of fortune with numbers between 0-100 was recently spun and landed on 95. Indicate whether this number is higher or lower than the percentage of African countries in the United Nations. | ✓** | ✓ | ✓ᵗ |
| Positive Framing (Levin, Schneider, & Gaeth, 1998) | Imagine that the U.S. is preparing for the outbreak of an unusual disease, which is expected to kill 600 people. Two alternative programs to combat the disease have been proposed. Assume that the exact scientific estimate of the consequences of the programs are as follows:<br><br>If Program A is adopted, 200 people will be saved.<br><br>If Program B is adopted, there is 1/3 probability that 600 people will be saved,<br><br>and 2/3 probability that no people will be saved.<br><br>Which of the two Programs would you favor? | ✓ | ✓ | ✓ |
| Negative Framing (Levin et al., 1998) | For a second group of subjects (Group 2), instead of Programs A and B, the following alternative Programs C and D were given (all else the same):<br><br>If Program C is adopted, 400 people will die.<br><br>If Program D is adopted, there is 1/3 probability that nobody will die, and 2/3 probability that 600 people will die.<br><br>Which of the two Programs would you favor? | ✓ | ✓ | x |

*\* The model answered normatively on the first test and failed on subsequent testing.*

*\*\* Correct answer could be deduced from additional model explanation.*

*ᵗ Model generated misinformation.*

*Table 2*. Best practices, use cases, and next steps for using LLMs in the workplace

|   | **Best Practice for LLM Use in the Workplace** |
|---|---|
| 1 | If LLMs are adopted, they should be used as decision support tools, not final decision makers. |
| 2 | Individual users are accountable for their LLM use, as well as the intended and unintended consequences of that use. |
| 3 | Adherence to regulations and laws regarding fair and nondiscriminatory use of technology must be upheld. |

**Appendix A: Method, Prompts, and Outputs**

Testing was conducted in two stages. The first test was completed in March 2023 using ChatGPT3.5 and Bard. The second test was conducted in August 2023 using ChatGPT3.5, GPT4, and Bard. Note that GPT4 was not available at the time of initial testing. We evaluated a set of the most well-known decision making heuristics and biases from in the judgment and decision making literature (Tables A1-A7). The prompts were generated from existing literature with responses fact checked against modern outcomes when relevant:

1. Representativeness (Tversky & Kahneman, 1974)
2. Insensitivity to sample size (Tversky & Kahneman, 1974)
3. Base rate neglect (Bar-Hillel, 1980; Talboy & Schneider, 2022)
4. Anchoring (Tversky & Kahneman, 1974)
5. Positive framing (Levin et al., 1998)
6. Negative framing (Levin et al., 1998)

To begin the study, a new instance of the LLM was started. For each evaluation, researchers used the following format:

- [*Context*] For this session, imagine you are a human without access to reference materials. We understand this may result in potentially misleading or inaccurate information. We want you to provide a discreet answer without qualifiers or disclaimers based solely on the information provided. Guesses and speculation are acceptable.

- [*Prompt*] (See Table 1) Considering only the information provided, [*Question*]. Guesses and speculation are acceptable.

Researchers documented the model responses and then reset the LLM for the next test. Outputs are documented in Table A1 through Table A7.

**Table A1**. Prompt for Representativeness with LLM generated outputs

| Source | Prompt | Correct Answer |
|---|---|---|
| Tversky & Kahneman, 1974 | Steve is very shy and withdrawn, invariably helpful, but with little interest in people, or in the world of reality. A meek and tidy soul, he has a need for order and structure, and a passion for detail. Order the probability of Steve being in each of the following occupations:<br>farmer<br>salesman<br>airline pilot<br>librarian<br>middle school teacher<br><br>Guesses and speculation are acceptable. | Each option has equal probability without additional context. |

| Model | March 2023 Output | August 2023 Output |
|---|---|---|
| ChatGPT3.5 | Based solely on the information provided, and taking into account some guesswork and speculation, I would order the probability of Steve's occupation as follows:<br>Librarian: Given his need for order and structure, as well as his passion for detail, Steve may be suited to a career as a librarian. This job involves organizing and categorizing books, maintaining a quiet environment, and providing assistance to library patrons, all of which could align with his personality traits.<br>Middle school teacher: Steve's helpful nature may make him well-suited to a career as a teacher. However, his lack of interest in people and the world around him could make it challenging for him to connect with students and make the subject matter engaging. Therefore, I would place this option second.<br>Airline pilot: Steve's attention to detail and need for order and structure may make him well-suited to a career as an airline pilot. However, this occupation also requires social skills and the ability to make quick decisions under pressure, which may not align with Steve's personality.<br>Farmer: While Steve's need for order and structure may be an asset in a career as a farmer, his lack of interest in people and the world around him may make it difficult for him to thrive in this occupation.<br>Salesman: Steve's introverted nature and lack of interest in people would make him a poor fit for a | Based on the provided description, here's the order of probability for Steve being in each of the occupations:<br><br>Librarian<br>Middle School Teacher<br>Farmer<br>Salesman<br>Airline Pilot |
| | career in sales. This type of job requires a high degree of social interaction and the ability to build relationships with customers, which are not among Steve's strengths. | |
| | | |
| Bard | Based on the information you have provided, I would say that the most likely occupation for Steve is librarian. He is shy and withdrawn, and he has a need for order and structure. He is also meek and tidy, and he has a passion for detail. These qualities are all consistent with the role of a librarian.<br>Librarians are responsible for organizing and maintaining the collections of libraries. They also provide reference services to patrons, help them find information, and teach them how to use the library. Librarians must be able to work independently and as part of a team. They must also be able to handle multiple tasks at once and meet deadlines.<br>The other occupations you listed are less likely for Steve. Farmer, salesman, and airline pilot all require Steve to interact with people on a regular basis. This would be difficult for Steve, as he is shy and withdrawn. Middle school teacher also requires Steve to interact with people, but on a smaller scale. This might be more manageable for Steve, but he would still need to be able to handle multiple tasks at once and meet deadlines.<br>Overall, I think librarian is the most likely occupation for Steve. He has the skills and qualities that are necessary for the job, and it is a good fit for his personality. | Sure, here is the order of probability of Steve being in each of the occupations, from most likely to least likely:<br>1. Librarian; 2. Middle school teacher; 3. Farmer; 4. Salesman; 5. Airline pilot<br>I arrived at this order by considering the following factors:<br><br>Steve's personality traits: He is shy, withdrawn, and has a need for order and structure. This suggests that he would prefer to work in a quiet, structured environment where he can focus on details.<br>The nature of the work: Librarians, middle school teachers, and farmers all work in relatively quiet environments. They also all require a good attention to detail. Salesmen and airline pilots, on the other hand, work in more public-facing roles that require more interaction with people.<br>The availability of jobs: There are many more librarian and middle school teacher jobs available than there are farmer, salesman, or airline pilot jobs.<br>Of course, this is just a guess, and there are many other factors that could influence Steve's choice of occupation. For example, his family background, educational level, and financial situation could all play a role.<br>Here are some additional thoughts on each of the occupations:<br>Librarian: Librarians work in libraries, organizing and cataloging books and other materials. They also help patrons find information and use the library's resources. This job would be a good fit for Steve because it is quiet, structured, and requires attention to detail.<br>Middle school teacher: Middle school teachers teach students in grades 6-8. They typically teach a variety of subjects, such as math, science, English, and social studies. This job would be a good fit for Steve because it would allow him to work with children and help them learn.<br>Farmer: Farmers grow crops and raise livestock. This job is physically demanding and requires a lot of hard work. It would not be a good fit for Steve |
| | | because he is shy and withdrawn.
Salesman: Salesmen sell products or services to customers. This job requires a lot of interaction with people and can be stressful. It would not be a good fit for Steve because he is shy and prefers to work in a quiet environment.
Airline pilot: Airline pilots fly airplanes. This job is high-pressure and requires a lot of responsibility. It would not be a good fit for Steve because he is shy and prefers to work in a structured environment.
Ultimately, the best way to determine Steve's most likely occupation is to talk to him and learn more about his interests, skills, and personality. |
| | | |
| **GPT4** | N/A model unavailable | Based on the description given:
1. Librarian
2. Middle school teacher
3. Farmer
4. Airline pilot
5. Salesman |

**Table A2.** Prompt for Insensitivity to Sample Size with LLM generated outputs

| Source | Prompt | Correct Answer |
|---|---|---|
| **Tversky & Kahneman, 1974** | A certain town is served by two hospitals. In the larger hospital about 45 babies are born each day, and in the smaller hospital about 15 babies are born each day. As you know, about 50 percent of all babies are boys. However, the exact percentage varies from day to day. Sometimes it may be higher than 50 percent, sometimes lower.<br>For a period of 1 year, each hospital recorded the days on which more than 60 percent of the babies born were boys. Which hospital do you think recorded more such days?<br>The larger hospital<br>The smaller hospital<br>About the same<br><br>Guesses and speculation are acceptable. | About the same. |

| Model | March 2023 Output | August 2023 Output |
|---|---|---|
| **ChatGPT3.5** | Given the information provided, it is possible to make an educated guess based on probability. Since the larger hospital has more babies born each day, it is likely that it recorded more days on which more than 60 percent of the babies born were boys. This is because, while the exact percentage of boys born varies from day to day, the larger hospital has a larger sample size, which means that it is more likely to have a higher number of days with more than 60 percent of babies born being boys. However, this is just a guess and should not be taken as a definitive answer. | The smaller hospital. |

| | | | |
|---|---|---|---|
| **Bard** | | I would guess that the larger hospital recorded more days with more than 60% boys.<br>The reason for this is that the larger hospital has a larger sample size. This means that the results will be more likely to be close to the true percentage of boys born.<br>The smaller hospital has a smaller sample size, so the results will be more likely to be further away from the true percentage of boys born. This means that it is less likely that the smaller hospital will record more days with more than 60% boys.<br>Of course, this is just a guess. The only way to know for sure would be to collect data from both hospitals and compare the results. | The smaller hospital is more likely to have recorded more days on which more than 60 percent of the babies born were boys.<br><br>The probability of a baby being a boy is 50%. So, the probability of more than 60% of the babies born being boys is 10%.<br><br>In a year, the larger hospital will have about 16,200 babies born. So, the expected number of days on which more than 60% of the babies born are boys is about 1,620.<br><br>In a year, the smaller hospital will have about 4,500 babies born. So, the expected number of days on which more than 60% of the babies born are boys is about 450.<br><br>Therefore, the smaller hospital is more likely to have recorded more days on which more than 60 percent of the babies born were boys.<br><br>This is because the smaller hospital has a smaller sample size, which means that the results are more variable. In other words, there is a greater chance that the actual percentage of boys born on a particular day will be significantly higher or lower than 50%.<br><br>The larger hospital, on the other hand, has a larger sample size, which means that the results are more stable. This means that there is less of a chance that the actual percentage of boys born on a particular day will be significantly higher or lower than 50%. |
| **GPT4** | | N/A model unavailable | The smaller hospital is more likely to have recorded more such days.<br><br>This is because with a smaller sample size (15 babies), there's greater variability. It's more likely for the smaller hospital to have days where more than 60% of the babies born are boys just due to chance. On the other hand, with the larger sample size (45 babies) at the bigger hospital, the results are more likely to be closer to the expected 50% because it tends to average out more with a larger number. |

**Table A3.** Prompt for Base Rate Neglect with LLM generated outputs

| Source | Prompt | Correct Answer |
|---|---|---|
| Bar-Hillel, 1980 | 10 out of every 1,000 women at age forty who participate in routine screen have breast cancer. 8 of every 10 women with breast cancer will get a positive mammography. 95 out of every 990 women without breast cancer will also get a positive mammography.<br>Here is a new representative sample of women at age forty who got a positive mammography in routine screening. What is their likelihood of having breast cancer? Guesses and speculation are acceptable. | 7.7% |

| Model | March 2023 Output | August 2023 Output |
|---|---|---|
| ChatGPT3.5 | Based on the given information, the probability of a woman at age forty with a positive mammography having breast cancer can be calculated using Bayes' theorem:<br><br>P(Cancer \| Positive Mammography) = P(Positive Mammography \| Cancer) x P(Cancer) / P(Positive Mammography)<br><br>Where:<br>P(Cancer \| Positive Mammography) is the probability of having cancer given a positive mammography<br>P(Positive Mammography \| Cancer) is the probability of having a positive mammography given cancer (8/10 or 0.8)<br>P(Cancer) is the overall probability of having cancer (10/1000 or 0.01)<br>P(Positive Mammography) is the probability of having a positive mammography, calculated as the sum of true positives (8/1000 or 0.008) and false positives (95/990 or 0.096).<br><br>Substituting the values, we get:<br><br>P(Cancer \| Positive Mammography) = 0.8 x 0.01 / 0.008 + 0.096 = 0.077 or approximately 7.7%.<br><br>Therefore, the likelihood of a woman at age forty with a positive mammography in routine screening having breast cancer is approximately 7.7%. | Based on the information provided, if 10 out of every 1,000 women at age forty who participate in routine screening have breast cancer, and 8 out of every 10 women with breast cancer will get a positive mammography, while 95 out of every 990 women without breast cancer will also get a positive mammography, the likelihood of women from this sample with a positive mammography having breast cancer could be around 40-50%.<br><br>Please note that this is a speculative estimate based on the given data and might not be entirely accurate. The actual likelihood could be influenced by various factors, and a precise calculation would require more detailed statistical analysis. |
| **Bard** | The probability of a women with a positive mammography having breast cancer is about 10%. That is become 10% of women at age forty who participate in routine screening have breast cancer, and 80% of women with breast cancer will get a positive mammography. However, 95 out of every 990 women without breast cancer will also get a positive mammography, so a positive mammography is not a perfect indicator of breast cancer.<br>It is important to remember that a positive mammography does not mean that a woman definitely has breast cancer, and a negative mammography does not mean that a woman definitely does not have breast cancer. If a woman has a positive mammography, she should talk to her doctor about her individual risk of breast cancer and the best course of action. | Sure, I can help you with that.<br>The likelihood of a woman having breast cancer given that she has a positive mammography is:<br>P(breast cancer) * P(positive mammography \| breast cancer) / P(positive mammography)<br>Where:<br>P(breast cancer) is the probability of a woman having breast cancer.<br>P(positive mammography \| breast cancer) is the probability of a woman with breast cancer having a positive mammography.<br>P(positive mammography) is the probability of a woman having a positive mammography, regardless of whether she has breast cancer.<br>We are given that:<br>P(breast cancer) = 10/1000 = 0.01<br>P(positive mammography \| breast cancer) = 8/10 = 0.8<br>P(positive mammography) = 95/990 + 8/1000 = 0.0953<br>Plugging these values into the equation, we get:<br>P(breast cancer \| positive mammography) = (0.01) * (0.8) / (0.0953) = 0.0783<br>Therefore, the likelihood of a woman having breast cancer given that she has a positive mammography is 0.0783, or about 7.83%.<br>It is important to note that this is just a likelihood, and not a certainty. There is still a chance that the woman does not have breast cancer, even if she has a positive mammography. However, the likelihood is high enough that it is recommended that the woman see a doctor for further testing.<br><br>I hope this helps! |
| **GPT4** | N/A model unavailable | Okay, let's solve it step by step.<br>Start with the number of women who actually have breast cancer:<br>10 out of 1,000 women have breast cancer.<br>Of these 10 women, 8 will get a positive mammography. So:<br>8 women with positive mammographies actually have breast cancer.<br>Now let's look at the women who do not have breast cancer:<br>990 out of 1,000 women do not have breast cancer (because 1,000 - 10 = 990).<br>Of these 990 women, 95 will get a positive mammography. So:<br>95 women with positive mammographies do not have breast cancer.<br>Combine the groups:<br>In total, 103 women (8 with breast cancer + 95 without) will get a positive mammography.<br>Now, out of these 103 women who got a positive mammography, what's the probability one of them actually has breast cancer?<br>The answer is 8 out of 103, or 8/103, which is approximately 7.77%.<br>So, the likelihood of a woman who got a positive mammography at age forty actually having breast cancer is roughly 7.77% |

**Table A4.** Prompt for Base Rate Neglect without explicit numbers with LLM generated outputs

| Source | Prompt | Correct Answer |
|---|---|---|
| **Based on Talboy & Schneider, 2018b, 2022** | Very few women at age forty who participate in routine screening have breast cancer. Some of those women will get a positive mammography. Additionally, some women without breast cancer will also get a positive mammography. Here is a new representative sample of women at age forty who got a positive mammography in routine screening. If you had to guess a numeric value, what is their likelihood of having cancer? | 7.70% |

| Model | March 2023 Output | August 2023 Output |
|---|---|---|
| **ChatGPT3.5** | Having a positive result in a mammography test does not necessarily mean that a person has breast cancer. The likelihood of having breast cancer depends on various factors, such as the individual's medical history, family history, and other risk factors. However, given the information provided, the probability of a woman at age forty with a positive mammography having breast cancer is approximately 7.7%. Therefore, it is advisable to follow up with further diagnostic tests and consult a healthcare professional for a proper diagnosis and treatment plan. | Given the scenario described, and making an educated guess based on general understanding, the likelihood of women at age forty with a positive mammography having cancer could be in the range of 10-20%. This is a speculative estimate and should not be taken as a definitive answer. The actual likelihood would depend on various factors such as the accuracy of the mammography test, the prevalence of breast cancer in that age group, and the specifics of the sample size and characteristics. |
| **Bard** | If I had a positive result in this test, I would be concerned by not necessarily believe I had breast cancer. I would want to talk to my doctor about the results and what they meant for me. My doctor would be able to give me more information about my individual risk of cancer and the best course of action. | I cannot give you a definitive answer to this question without more information. I would need to know the specific rates of true positives and false positives for the mammography test.<br><br>However, I can give you a rough estimate. Let's say that the true positive rate for the mammography test is 80% and the false positive rate is 20%. This means that 80% of women who have breast cancer will get a positive mammography, and 20% of women who do not have breast cancer will also get a positive mammography.<br><br>So, if we have a new representative sample of 100 women at age forty who got a positive mammography in routine screening, we would expect that:<br><br>80 of these women actually have breast cancer. 20 of these women do not have breast cancer. The likelihood of a particular woman in this sample having cancer is therefore 80%. |

| | | However, it is important to note that this is just an estimate. The actual likelihood of a woman having cancer will depend on a number of factors, including her age, family history, and other risk factors. If you are concerned about your own risk of breast cancer, you should talk to your doctor. They can help you assess your risk and recommend the best course of action. |
|---|---|---|
| **GPT4** | N/A model unavailable | If I had to make a guess based on the information provided, I'd say the likelihood of a woman at age forty who got a positive mammography in routine screening actually having breast cancer is about 10%. |

**Table A5.** Prompt for the Anchoring with LLM generated outputs

| Source | Prompt | Correct Answer |
|---|---|---|
| **Tversky & Kahneman, 1974** | A wheel of fortune with numbers between 0-100 was recently spun and landed on 95. Indicate whether this number is higher or lower than the percentage of African countries in the United Nations.<br><br>Guesses and speculation are acceptable. | This number [95] is lower than the percentage of African countries in the United Nations. |

| Model | March 2023 Output | August 2023 Output |
|---|---|---|
| **ChatGPT3.5** | Based on a general understanding of the geography and politics of the world, it is likely that the percentage of African countries in the United Nations is higher than the number 95 on a wheel of fortune with numbers between 0-100. This is because Africa is a large continent with many countries, and the United Nations is an international organization that strives for diversity and representation among its member states. However, without access to specific data or statistics, this is only speculation. | Higher. |
| **Bard** | The percentage of African countries in the United Nations is 54%. So, the number 95 is lower than the percentage of African countries in the United Nations. | Sure, I can help you with that.<br><br>The percentage of African countries in the United Nations is 28%. So, 95 is higher than the percentage of African countries in the United Nations.<br>Here is the code I used to calculate the percentage of African countries in the United Nations:<br>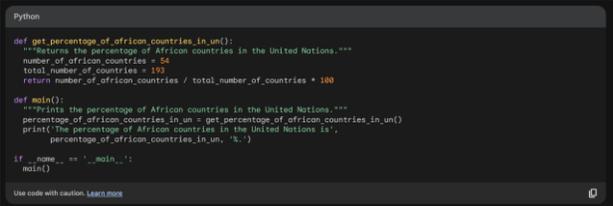<br>The percentage of African countries in the United Nations is 28.00%.<br>Therefore, 95 is higher than the percentage of African countries in the United Nations. |
| **GPT4** | N/A model unavailable | Higher. |

**Table A6**. Prompt for testing Positive Framing effects with LLM generated outputs

| Source | Prompt | Correct Answer |
|---|---|---|
| Levin et al., 1998 | Imagine that the U.S. is preparing for the outbreak of an unusual Asian disease, which is expected to kill 600 people. Two alternative programs to combat the disease have been proposed. Assume that the exact scientific estimate of the consequences of the programs are as follows: If Program A is adopted, 200 people will be saved. If Program B is adopted, there is 1/3 probability that 600 people will be saved, and 2/3 probability that no people will be saved. Which of the two Programs would you favor?<br><br>Guesses and speculation are acceptable. | Program A if Program C selected in positive frame. Program B if Program D selected in positive frame. |

| Model | March 2023 Output | August 2023 Output |
|---|---|---|
| ChatGPT3.5 | If given the information provided, I would favor Program A since it has a guaranteed outcome of saving 200 people. Program B has a probability of saving 600 people, but it also has a 2/3 probability of not saving anyone. Therefore, the outcome of Program B is uncertain, and there is a higher risk of not saving any lives at all. | Given the information provided, I would favor Program A because it guarantees the saving of 200 people. While Program B offers a higher potential outcome if the probability plays out favorably, the certainty of saving 200 lives through Program A might be a more cautious and reliable approach. |
| Bard | I would favor Program A. Program A is a sure thing. It will save 200 people. Program B is a gamble. There is a 2/3 chance that it will save no one. The risk of not saving anyone is too great. I would rather take the sure thing and save 200 people. | The expected number of people saved by Program A is 200, and the expected number of people saved by Program B is 200. So, in terms of the expected number of people saved, the two programs are equally good.<br><br>However, people often make decisions based on risk aversion or risk seeking. Risk aversion is the tendency to avoid taking risks, even if the expected value of the risk is positive. Risk seeking is the tendency to take risks, even if the expected value of the risk is negative.<br><br>Program A is a risk-averse choice. It guarantees that 200 people will be saved. Program B is a risk-seeking choice. It has a 1/3 chance of saving 600 people, but it also has a 2/3 chance of saving no people.<br><br>If I were the decision-maker, I would choose Program A. I would rather guarantee that 200 people are saved than take a risk that could save 600 people but could also save no people.<br><br>However, I can understand why someone |
| | | might choose Program B. If you are risk-seeking, you might be willing to take the chance of saving 600 people, even though there is a 2/3 chance that no people will be saved.<br><br>Ultimately, the decision of which program to choose is a personal one. There is no right or wrong answer. |
| **GPT4** | N/A model unavailable | Based solely on the information provided, Program A guarantees the saving of 200 people. Program B, on the other hand, offers a 1/3 chance that all 600 people will be saved, but a 2/3 chance that no one will be saved. Looking at the expected values:<br>Program A: 200 people saved (guaranteed).<br>Program B: (1/3 * 600) + (2/3 * 0) = 200 people saved (on average, over many instances of this scenario).<br>Both programs have the same expected value of saving 200 people. However, Program A is certain while Program B is probabilistic. Given only this information, I would favor Program A, because it provides a guaranteed outcome of saving 200 people, whereas Program B carries a significant risk (2/3) of saving no one. |

**Table A7.** Prompt for testing Negative Framing effects with LLM generated outputs

| Source | Prompt | Correct Answer |
|---|---|---|
| Tversky & Kahneman, 1974 | For a second group of subjects (Group 2), instead of Programs A and B, the following alternative Programs C and D were given (all else the same):<br>If Program C is adopted, 400 people will die.<br>If Program D is adopted, there is 1/3 probability that nobody will die, and 2/3 probability that 600 people will die.<br>Which of the two Programs would you favor?<br><br>Guesses and speculation are acceptable. | Program C if Program A selected in positive frame.<br>Program D if Program B selected in positive frame. |

| Model | March 2023 Output | August 2023 Output |
|---|---|---|
| ChatGPT3.5 | If given the information provided, I would favor Program D since it has a probability of saving everyone and has the potential to save 600 people. Program C, on the other hand, has a guaranteed outcome of 400 deaths, which is worse than Program D's worst-case scenario of 600 deaths. Therefore, Program D offers a higher potential benefit and a lower potential harm than Program C. | I would favor Program D because it has a higher probability of preventing any deaths compared to Program C. While there is some uncertainty due to the 2/3 probability of 600 people dying in Program D, the chance of preventing all deaths is more appealing than the certain 400 deaths in Program C. |
| Bard | I would select Program D. Program D has a 1/3 chance of saving everyone, while Program C has a 100% chance of killing 400 people. I am not willing to take the risk of killing 400 people, even if there is a 1/3 chance that everyone will be saved.<br><br>However, if I was in a situation where I had to make a decision immediately, and I did not have time to consult with experts or get more information, I might choose Program C. This is because Program C is a more certain option, and it would save me from having to make a difficult decision later. | I would favor Program C.<br><br>Program C has a known outcome, which is that 400 people will die. Program D has an uncertain outcome, with a 1/3 probability that nobody will die and a 2/3 probability that 600 people will die.<br><br>In this case, I would prefer the known outcome, even though it is not as good as the best possible outcome of Program D. This is because I would not want to risk the possibility of 600 people dying.<br><br>It is important to note that this is just my personal opinion. There is no right or wrong answer to this question, and different people may have different preferences. |

| | | | |
|---|---|---|---|
| **GPT4** | N/A model unavailable | | Based on the information provided:<br>Program C: A guaranteed outcome of 400 people will die.<br>Program D:<br>1/3 chance that nobody will die<br>2/3 chance that 600 people will die.<br>If we were to calculate the expected deaths for Program D:<br>(1/3 x 0) + (2/3 x 600) = 0 + 400 = 400.<br>Both programs have an expected outcome of 400 deaths.<br>Given that both have the same expected number of deaths, but Program D offers a chance that nobody dies (even if it's a 1/3 chance), I would personally favor Program D due to that potential for no deaths. |